\documentclass[fleqn,usenatbib,letters]{mnras}

\usepackage[T1]{fontenc}
\usepackage{ae,aecompl}
\usepackage{xcolor}

\usepackage{graphicx}	% Including figure files
\usepackage{amsmath}	% Advanced maths commands
\usepackage{amssymb}	% Extra maths symbols

\title[ORCs as intragroup SNRs]{Odd Radio Circles as supernovae remnants in the intragroup medium}

\author[A. Omar]{
A. Omar\thanks{E-mail: aomar@aries.res.in (AO)}
\\
Aryabhatta Research Institute of observational-sciences, Manora Peak, Nainital, 263001, India\\
}

\begin{document}
%\label{firstpage}
\maketitle

\begin{abstract}
A measurable fraction ($\sim8$ per cent) of recently discovered arcmin-size circular diffuse radio sources termed as Odd Radio Circles or ORCs can be supernovae remnants in the intragroup medium, within the local group and its immediate neighbour groups of galaxies. This estimate is based on the optical detection rate of the intragroup supernovae events in the nearby ($z \sim 0.1-0.2$) galaxy groups. A rate of about 5400 intragroup supernovae per million year is expected within the local and its immediate neighbour groups of galaxies. For a radio detectability period of about $10^{4}$ years, on average 1.3 intragroup medium supernovae remnants per 1000 square degree are expected to be detected in the radio surveys with a sensitivity that led to discovery of ORCs. The angular size, surface brightness and radio flux of the supernova remnants up to a distance of $\sim3$ Mpc in the intragroup medium can be expected to be similar to the five known ORCs. The intragroup supernovae remnants are not residing in the dense and cold interstellar medium of the galaxies but evolving in low density ($10^{-4}-10^{-5}$ cm$^{-3}$) warm medium ($10^{5}-10^{6}$ K) in galactic halos or beyond, and may find their progenitors in the diffuse stellar light associated with various tidal streamers surrounding the Milky-Way and other nearby galaxies.  

\end{abstract}


%%%%%%%%%%%%%%%%%%%%%%%%%%%%%%%%%%%%%%%%%%%%%%%%%%

%%%%%%%%%%%%%%%%% BODY OF PAPER %%%%%%%%%%%%%%%%%%

\section{Introduction}

Some five Odd Radio Circles (ORCs) were recently discovered as a new population of extremely faint (surface brightness a few tens of $\mu$Jy per beam) diffuse radio sources in the celestial sky, all having circular ring-like morphology with nearly identical angular diameters of $\sim1$ arcmin \citep{Nora, Norc, Kori}. These sources were identified either in a deep pilot radio survey named as Evolutionary Map of the Universe (EMU; Norris et al. 2021b),  made using the Australian Square Kilometre Array Pathfinder (ASKAP) or in the Giant Meter-wave Radio Telescope (GMRT) archival data. The EMU survey attained surface brightness sensitivity of $\sim30$~$\mu$Jy per beam rms at an angular resolution of nearly 12 arcsec near 1 GHz (0.944 GHz) frequency \citep{Norb}. Three of these ORCs were found in a search area of 270 deg$^{2}$ of the ASKAP data, one in 40 deg$^{2}$ area around the galaxy NGC~253 in the ASKAP data, and one in the archival observation data at 325~MHz in the cluster field Abell 2142 from the GMRT. The EMU data cover fields around the Galactic latitude of $-40$ degree. These radio objects are named as ORC-1 to ORC-5 with their Galactic $(l,b)$ coordinates at (333.42, -39.0), (339.09, -39.5), (339.08, -39.6), (44.36, 49.4) and (170.78, -86.6) degree respectively. 

The morphological appearances of the five ORCs are strikingly similar to a typical Galactic supernovae remnant (SNR). However, \citet{Nora} ruled out these as Galactic SNRs based on the high latitude positions of the ORCs coming at odd with the statistical distribution of the known SNRs as a function of the Galactic latitude. Although, a possibility for ORCs as a previously unknown population of SNRs was not ruled out. All the five ORCs have a similar angular diameter of nearly 1 arcmin in the sky. Optical counterparts have been associated near the centers of ORC-1, ORC-4 and ORC-5 at photometric redshifts of 0.551, 0.457 and 0.270 respectively. ORC-2 and ORC-3 could not be associated with any optical counterparts. The ORC-2 and ORC-3 also form a close pair with an angular separation of nearly $1.5'$ in the sky. The flux densities near 1 GHz for four of these ORCs are almost similar at $\sim7$ mJy. The ORC-3 is the faintest with flux density at $\sim2$ mJy. The mean radio surface brightness of 4 bright ORCs is nearly $10^{-21}$ W~m$^{-2}$~Hz$^{-1}$~sr$^{-1}$ and that of the faintest ORC is nearly $3\times10^{-22}$ W~m$^{-2}$~Hz$^{-1}$~sr$^{-1}$. Their radio spectral indices are negative, indicating that the radio emission is of non-thermal origin. 

The astrophysical origins of the ORCs are presently debated with a number of possibilities. \citet{Nora} discussed various possibilities to associate ORCs with different types of known astrophysical sources after excluding a possibility of Galactic SNRs. However, majority of other possibilities are also ruled out on the basis of ORC's distinct morphological features such as edge-brightening, symmetry and a high degree of circularity, which are not seen in other known similar diffuse radio sources such as lobes of radio galaxies, face-on star-forming galaxies and cluster halos. Moreover, ORCs are not physically associated with large face-on galaxies or rich galaxy clusters. \citet{Norb} and \citet {Kori} mentioned three promising theoretical possibilities for the origins of some of the ORCs, viz., spherical shock resulting from a merger of two supermassive black-holes, vortex rings in radio lobes viewed end-on in radio galaxies, and termination shock of starburst winds from the galaxy. Presently, any single or all five ORCs can not be firmly associated with a definite type of astrophysical object or phenomena.

In this paper, we discuss a possibility that some ORCs may trace their origins in SNRs in low density intragroup medium (IGrM). We will use single abbreviation IGrM to denote both intragroup medium commonly associated with the local group, and intergalactic medium or IGM commonly used for other groups of galaxies. The motivation for exploring this possibility comes from detections of a subtle population of stars in the IGrM in the form of various known stellar streams in our local group and in other nearby groups of galaxies, diffuse intracluster light in some clusters of galaxies, and detections of a significant number of host-less supernovae. 

\section{ORC\lowercase{s} as SNR\lowercase{s} in low density hot medium}

Based on a similar appearance as those of the Galactic SNRs, the ORCs are very likely dynamically evolving objects, created via a catastrophic event in some astrophysical objects. As the interest here is to explore a possibility of some ORCs being SNR, and considering that the ORCs do not appear to be residing in the Milky-Way or in any other large galaxy, a discussion of evolution of SNRs in low-density hot medium, typical in IGrM environments will be the most relevant here. 

The radio brightness and morphology of a SNR depends on various factor such as age, type (I or II) based on the properties of the progenitor star, and density and temperature of the ambient medium in which it is evolving. Early numerical studies related to surface brightness and size evolution of SNRs as a function of ambient temperature and gas density were carried out by \citet{Tom} and \citet{Hig}. In their minimalistic theoretical framework, the sizes of SNRs are expected to become larger and surface brightness to become fainter in low-density warm medium (e.g., $n\sim10^{-4}$~cm$^{-3}$, $T\sim10^{5}$ K), compared to their counterparts in dense and cold (e.g., $n\sim10^{-1}$~cm$^{-3}$, $T\sim10^{2}$ K) interstellar medium (ISM) in the Galactic plane. Evolutions of SNRs in hot medium with a temperature up to $5\times10^{7}$ K are described in \citet{Tang}. Based on detailed hydrodynamical simulations, estimates for size and radio surface brightness variations of SNRs with ambient density and time are provided in more recent work of \citet{pavl}. The main results from all these studies can be summarized as follows -- (i) Expansion deviates quickly ($<10^{4}$ years) from the standard Sedov-Taylor solutions at ambient temperatures above $10^{6}$ K. (ii) The sizes of SNRs near the end of the Sedov-Taylor phase are expected to scale inversely with ambient density as a power-law with an exponent of $\sim1/3$, (iii) The SNRs expanding in a hot medium are expected to have a lower radio surface brightness than those in a warm medium. This behaviour is due to lower Mach number of the shock in a hotter medium. (iv) The radio surface brightness decreases very fast in late phases (i.e., Sedov-Taylor regime) of evolution with surface brightness to diameter slopes in between $-4$ and $-6$. It means that while the diameter will not increase very fast (i.e. expansion is slowed down), the surface brightness decreases very fast in the late stages. These general results will also be used in the next section to obtain a range of possible distances to ORCs.  

It is also worth to mention here that the shell component of SNRs in a low ambient density will brighten at late stages and are likely to be detected in radio surveys, compared to the plerionic component which will be bright in the beginning for only a very short duration and will become extremely faint after a few 100 years from the explosion (e.g., \citealt{DB1}). The dynamical lifetime of SNRs are expected to be $10^{4}-10^{5}$ years, after which SNRs mix with the general ISM and cease to exist as a morphologically distinct structure. It may be noted that the synchrotron lifetime of relativistic electrons generated in a SNR are expected to be much higher (up to $\sim100$ Myr) in low (a few $\mu$G) magnetic field strengths, however, these electrons will diffuse out to much larger extents during this period and will contribute to large-scale diffuse radio emission. This simple dynamical behaviour of SNRs will have two consequences - firstly, small size SNRs representative of younger age will be difficult to detect in a survey covering low volume in the space and secondly, beyond a certain stage of expansion at later times, SNRs will become too faint and large to be detected in a survey with a limited sensitivity. Hence, statistically, if some ORCs are SNRs, then these are detected at a particular stage of their evolution, when both surface brightness and angular diameter become favourable for their detections in the radio survey with a limited sensitivity. This stage will be neither too early nor too late.

It is a well established fact that a large number of Galactic SNRs are missing in the present radio surveys due to multiple reasons (see e.g., \citealt{DB1, green}). The main reason is believed to be that many SNRs evolve in low-density hot medium, created in the cavities blown by the progenitor massive star thereby making such SNRs large and faint. The biases created due to sensitivity limits in any radio surveys are also the reasons for missing some SNRs. For instance, sensitivities of the radio interferometric surveys such as ASKAP are limited in both flux and in detectable largest angular diameter, which is decided by the shortest distance in the pairs of the antennas in the radio array. Due to increase in sensitivity of the present generation radio telescopes and with the advent of new detection techniques, some very large and very faint SNRs have recently been detected. For instance, a large size ($\sim4.4$ degree diameter) SNR named 'Hoinga (G249.7+24.7)' with very low radio surface brightness ($4\times10^{-23}$ W~m$^{-2}$~Hz$^{-1}$~sr$^{-1}$ at 1.4 GHz) was detected at Galactic latitude of $\sim25$ degree \citep{Beck, fes}. Another SNR 'G354-33' with angular size of $11^{0}\times14^{0}$ at Galactic latitude of $-33.5$ degree was detected in radio \citep{fes}. Detections of such large SNRs are obviously more favourable at higher Galactic latitudes due to reduced Galactic diffuse radio background at high latitudes. 

Since ORCs have low radio surface brightness and have been detected due to a breakthrough in detection sensitivity with the ASKAP, it is worth to explore their origins in low-density hot medium SNRs, which are believed to be missing in previous generation radio surveys. The biggest challenge will be to constrain the distances as discussed in the next section.  

\section{Distances to ORC\lowercase{s}}

It becomes worth at this point to compare properties of the ORCs with those of the Galactic SNRs and obtain some clues about their distances. Majority of the Galactic radio SNRs have their 1 GHz radio luminosity in the range of $10^{15} - 10^{18}$ W Hz$^{-1}$ and radio surface brightness in the range of $10^{-22} - 10^{-18}$ W~m$^{-2}$~Hz$^{-1}$~sr$^{-1}$ \citep{green}. The average radio surface brightness of five ORCs is in between $10^{-22} - 10^{-21}$ W~m$^{-2}$~Hz$^{-1}$~sr$^{-1}$, placing ORCs among the faintest bins of radio surface brightness of the Galactic SNRs. 

Considering ORCs as SNRs at late stages where their expansion is slowed down, their angular sizes can provide another clue to their distances. The results provided in \citet{pavl} for SNRs evolving in ambient medium temperature of $10^4$ K and a range of densities between 2 to 0.005 cm$^{-3}$ for initial explosion energy in the range of 0.5 to 2.0$\times10^{51}$ erg can be adjusted for lower densities and higher temperatures in the IGrM, using the general results described in the previous section. In ambient density $10^{-3}$ cm$^{-3}$, the 1 GHz radio surface brightness between $10^{-21}$ and $10^{-22}$ W~m$^{-2}$~Hz$^{-1}$~sr$^{-1}$ will correspond to SNR diameter of $\sim100$ pc. In lower ambient density, e.g., in $10^{-4}$ cm$^{-3}$ and $10^{-5}$ cm$^{-3}$, the sizes can be up to $\sim200$ pc and $\sim500$ pc respectively. It may be noted that the effect of increased ambient temperature will be on the radio surface brightness. The SNRs in hot (T$>10^{6}$~K) ambient medium (e.g., in galaxy clusters or massive groups with diffuse X-ray emission) are expected to become fainter than those in relatively warm (T$\sim10^{5}-10^{6}$~K) ambient medium, typically seen in low mass groups of galaxies. This effect can reduce chances of detecting as well as identifying (from a background un-resolved radio source population) SNRs in the cores of galaxy clusters and in groups with diffuse X-ray emission.

Considering an upper limit to physical diameter as 500 pc corresponding to the detected range of radio surface brightness of the ORCs, the maximum distance to the  ORCs in a low density ($10^{-5}$ cm$^{-3}$) warm ($10^{5}$K) IGrM environment can be about 1.5 Mpc for an average angular diameter of 1 arcmin. If ORCs are evolving in higher density (e.g., $10^{-3}$ cm$^{-3}$) ambient medium, this distance can be revised down to 0.3 Mpc. The 1 GHz radio luminosity of the Galactic radio SNRs are generally in the range of $10^{15} - 10^{18}$ W Hz$^{-1}$ \citep{green}. Assuming that the radio luminosities of the ORCs are similar to that of the general population of the Galactic SNRs, the ORCs with their estimated flux density can be placed in a distance bracket within a range of $\sim0.1$ Mpc to $\sim3$ Mpc. Therefore, the range of estimated distances of the ORCs is nearly 0.1 - 3 Mpc and the two arguments, viz., using radio brightness and size, and radio luminosity fetch similar results within the uncertainties. 

\section{Discussions}

\subsection{On whereabouts of ORCs}

From the arguments and estimates for the distances presented in the previous section, the SNRs in IGrM appearing as ORCs can be best placed within the extent of the local group of galaxies or some of its immediate neighbours (see \citealt{van} for properties of the nearby groups). The radius of the local group is estimated as $\sim1.2$ Mpc. More than 35 galaxies are currently designated as members of the local group with a few additional probable candidates and several other galaxies placed in the immediate neighbour groups within a distance of $\sim5$ Mpc. The locations of the galaxies in the local group are not uniformly distributed in the sky. As no bright galaxy from the local or its neighbour groups is detected at the locations of the ORCs, some of the ORCs can be predicted to be residing in the IGrM. The IGrM of the local group is not completely empty of stars. More than 40 stellar streams have been discovered at a range of distances surrounding the Milky Way \citep{grill}. These streams originated during the tidal interactions between galaxies in the group. In general, the environment of a typical galaxy group is favourable to exert significant tidal disturbances on galaxies, leading to removal of gas and stars from the parent galaxies to much larger extents in the IGrM (e.g., \citealt{omar2}). Two largest stellar streams in the local group are associated with Sagittarius dwarf galaxy and Large Magellanic Cloud (LMC) - Small Magellanic Cloud (SMC) at a distance of nearly 60 kpc. The Sagittarius stream is seen almost over the entire sky. The LMC-SMC stream is very wide and data from $Gaia$ suggest that the stars are observed around LMC to an extent up to $20$ degree \citep{belo}. 

The stellar streams are populated mostly by low-mass stars, which may undergo type Ia supernovae (SNIa). Some ORCs may be SNRs in one of these stellar streams. It is worth to note here that three ORCs are detected in a area surrounding SMC and one near NGC 253 - a sculptor group galaxy at a distance of $\sim3$ Mpc.  ORC-1, ORC-2 and ORC-3 are within 23-27 deg (about 30 kpc) from the SMC. ORC-5 is at an angular separation of about 3 deg (about 170 kpc) from NGC 253. The estimated separations of ORCs from the SMC or NGC 253 are not large that the ORCs can not be associated with these galaxies. A number of host-less (i.e., not associated with a galaxy) supernovae in the IGrM have been detected in optical bands \citep{Gal-yam, Sand, Dilday, sand11}. \citet{Mcgee} provided a statistics of the host-less SNIa detection rate in groups of galaxies based on total supernovae detected in the Sloan digital sky supernovae survey \citep{frie} after assigning memberships of the detected supernovae across 1401 groups of galaxies in the redshift range $0.1 - 0.2$. A total of 22 SNIa were found to be host-less in the galaxy groups, in the total survey-observing period of nearly 9 months. The combined halo mass of all the groups was estimated to be $5.4\times10^{16}$~M$_{\odot}$. It translates in to an average of 5400 IGrM or host-less supernovae per million year in a group mass of $10^{13}$~M$_{\odot}$. 

The total mass of the local group is estimated as $2.3\times10^{12}$ M$_{\odot}$ \citep{van}. Assuming that the masses of neighbour groups, namely Antlia-Sextans group at 1.7 Mpc, the Sculptor group at 2.4 Mpc, the IC 342/Maffei group at 3.2 Mpc and the M 81 group at 3.5 Mpc are of similar order, total mass of the local group and its immediate neighbours can be taken as nearly $10^{13}$ M$_{\odot}$. Therefore, the SNIa rate in the IGrM associated with the local group and its immediate neighbour groups is about 5400 supernovae per million year. If the remnants of these supernovae are detectable up to $10^{4}$ years in the radio bands, about 54 such SNRs may be detected in the sky with a corresponding 1 GHz radio surface brightness between $10^{-22} - 10^{-21}$ W~m$^{-2}$~Hz$^{-1}$~sr$^{-1}$. This estimate can be significantly uncertain as it depends on the assumed density and temperature of the IGrM, which is poorly constrained. However, this estimate is viable and can be considered as an upper limit as long as the mean temperature of the IGrM does not exceed $10^{6}$ K (see discussions in \citealt{mul} for IGrM temperature in the local group). If IGrM temperatures are lower, SNRs detectability may increase significantly for the older ($>10^{4}$ years) supernovae. 

The number of progenitors or the intragroup SNIa rate may also differ from one group to other as it can depend on the dynamical stage of the group and its past evolutionary history in terms of tidal interactions and mergers. Nevertheless, the average SNIa rate estimated from a large number (1401) of low-redshift groups covered in the survey has a high statistical significance and can be taken as a most probable average rate among the low redshift galaxy groups. Therefore, a statistically significant possibility to detect some SNRs in the IGrM of the local group and its immediate neighbour groups of galaxies with ASKAP exists with a detection rate of about 1.3 SNR per 1000 square degree. As the present detection rate of ORCs is much higher at about 1.6 every 100 square degree, not more than about 8 per cent of ORCs can be associated with SNRs in the IGrM of local groups and its immediate neighbour groups of galaxies. This detection rate is likely to change, as more ORCs will be detected in future surveys covering a much larger area in the sky. The main outcome of this analysis is that the expected detection rate of local group SNRs in the IGrM is non-negligible yet much smaller than the present rate of detections of ORCs. A similar analysis for the detectability of intergalactic SNRs in the Virgo group of galaxies was previously presented in \citet{Maoz05}. They estimated about 10 SNRs per square degree with 1 GHz radio flux of 0.1 mJy in Virgo galaxy cluster at an assumed distance of 10 Mpc and for SNR detectability lasting for $10^{4}$ years. 

\subsection{On sizes and evolutionary stages of ORCs}

Two most intriguing features of the ORCs are nearly identical angular size ($\sim1$ arcmin) for all the sources and similar radio brightness for 4 out of 5 sources. If ORCs are dynamically evolving objects similar to SNRs, a distribution in size and flux density is expected in a survey. There could be multiple reasons for not seeing this distribution. The ASKAP being an interferometric survey loses sensitivity to detect extended sources beyond about $\sim1$ arcmin \citep{Norb}. The convolved angular resolution ($\sim20$ arcsec) of the ASKAP images may not be sufficient to distinguish ORCs smaller than $\sim1$ arcmin from the known population of other diffuse radio sources. If the ORCs are residing at widely different redshifts ($z\sim0.2 - 0.6$), the physical diameters of the distant ORCs will be larger than those of the nearby ORCs, but both the cases having identical surface brightness. If ORCs are dynamically evolving objects similar to SNRs, the surface brightness is expected to reduce with increasing physical diameter. Here, the similarity with a SNR is drawn in the context of dynamical behaviour of an expanding shock and not in terms of input mechanical energy, which could be much higher in case of super massive black hole merger scenario invoked to explain ORCs in distant galaxies by \citet{Norc} and \citet{Kori}. On the other hand, if ORCs are nearly equidistant, e.g., in some nearby structure or a group of galaxies covering large angular extent in the sky, the variations in sizes and brightness are expected to be less. Therefore, physical origins of the ORCs presently associated with distant galaxies needs to be explained in a theoretical framework, which allows a substantial variation of the key dynamical parameters such as input mechanical energy from the initial explosion, and density and temperature of the ambient medium. As ORCs seem to belong to different types of objects, their similar appearances are likely to be related to dynamical evolution of expanding shocks but with different energies in different astrophysical processes. 

The detection of two ORCs namely ORC-2 and ORC-3 at a very close separation in the sky possibly also offers another clue about their evolutionary stages. These two ORCs are separated by only about 1.5 arcmin in the sky. Considering the rarity of ORCs, these two ORCs can be expected to be physically nearby in space hence nearly equidistant and are not just a chance superposition in the sky of two widely separated objects. As no optical counterparts have been associated with these two ORCs, these may be genuine SNRs in the IGrM. These two ORCs have nearly equal angular diameter but different flux,  $\sim7$ mJy for ORC-2 and $\sim2$ mJy for ORC-3. If these two ORCs are SNRs co-evolving in an identical ambient environment and at a same distance from us, their identical angular sizes may imply that these SNRs are seen at a late stage of evolution where the expansion of the shock has already slowed down. The difference in flux is then not due to a distance variation but due to their age difference where the synchrotron losses makes the older object appear fainter. The ORC-2 and ORC-3 can then be SNRs detected in the IGrM in a late stage of evolution, when expansions of the SNRs have already slowed down. The mystery still remains that how two such SNRs come so close in the space. 
 
\section{Summary and concluding remarks}

In this paper, radio surface brightness, luminosity, and angular size of the ORCs were discussed. We attempted here to trace origins of some ORCs in the SNRs in the IGrM. Our estimates made under a minimalistic framework, suggest that some ORCs can be SNRs in the IGrM of the local group and in its immediate neighbour groups of galaxies within a distance of about 3 Mpc. The estimated host-less (presumably in IGrM) optical supernovae rate in other nearby groups suggests that only about 1.3 supernovae per 1000 square degree may be detected, assuming a detectability in radio for about $10^{4}$ years. The detectability of old SNRs at an epoch may depend upon the temperature of the IGrM in the sense that in hot IGrM (T$>10^{6}$ K), radio bright phase of the SNRs will last for a shorter period than those in a relatively warmer IGrM. This effect may favour larger number of SNR detections in the galaxy groups than in the galaxy clusters. As the present detection rate of ORCs is much higher at about 1.6 every 100 square degree, the ORCs as a class of objects will be heterogeneous in nature and will very likely trace their origins in multiple astrophysical objects or events. A similarity in physical appearances of five ORCs indicates that although the ORCs may trace their origins in multiple types of objects or events, the underlying phenomena responsible for their morphologies is likely to be the expanding shock waves in IGrM. With the present detection rate estimated from just five ORCs, not more than about 8 per cent of ORCs may be associated as SNRs in the IGrM of the local group and its immediate neighbours. This value is likely to change, as more ORCs will be detected in future. 

Out of the five known ORCs, three ORCs have been associated with an optical counterpart and two ORCs do not have optical counterparts. Presently, possible explanations of the origins of ORCs do not seamlessly fit into a single previously observed energetic phenomena. As some ORCs have an optical galaxy and some appear host-less, finding origins of ORCs in multiple types of physical objects and processes appears a clever choice. Given that the present sample of ORCs is statistically very small and so any extrapolation of one possible explanation to all such objects need not be carried out and at the same time any possible or viable scenario should not be left out. Presently, these important questions remain to be answered - (1) The nearly identical sizes and radio flux of ORCs associated with an optical galaxy pose a challenge to understand their origins in deep extra-galactic objects at widely different redshifts. (2) Two ORCs with a very close separation in the sky. (3) Missing smaller angular size ORCs in the ASKAP survey. 

Finally, it is worth to mention here some important consequences of confirming detections of SNRs in the IGrM. 
\citet{Domainko04} and \citet{Zaritsky} discussed importance of intracluster supernovae to match the observed Fe abundance excess in the cool-core clusters. Supernovae are also considered efficient sources of heating and turbulence over large volumes in low-density hot medium typical of that in the intracluster medium \citep{Valdarnini, Domainko04, Tang}. \citet{omar} discussed a possibility that radio emission from mini halos, another low radio surface brightness structures on physical scales of a few hundred kpc, undoubtedly associated with the brightest cluster galaxies in cool-core galaxy clusters, can be explained in terms of relativistic electrons accelerated in the intracluster supernovae events. A direct confirmed detection of intergalactic SNRs will strengthen this possibility. The unprecedented sensitivity combined with the anticipated sub-arcsec resolution of the upcoming Square Kilometer Array telescope may enable us to identify SNRs in the IGrM of some nearby galaxy groups and clusters.

% Don't change these lines
\bsp	% typesetting comment
\label{lastpage}
\end{document}